\documentclass{article}
\usepackage{spconf,amsmath,graphicx}
\usepackage{multirow}
\usepackage{subfigure}
\usepackage{color}
\usepackage{xspace}
\usepackage{hyperref}
\usepackage{amssymb}

\newcommand{\ie}{{\em i.e.,}\xspace}

\title{Contextual Expressive Text-to-Speech}

\name{
Jianhong Tu\textsuperscript{1,2,*}, 
Zeyu Cui\textsuperscript{1,*}\thanks{* Equal contribution}, 
Xiaohuan Zhou\textsuperscript{1}, 
Siqi Zheng\textsuperscript{1}, 
Kai Hu\textsuperscript{1}, 
Ju Fan\textsuperscript{2}, 
Chang Zhou\textsuperscript{1,$\dagger$}\thanks{$\dagger$ Corresponding author}}
\address{Alibaba DAMO Academy\textsuperscript{1}, Renmin University of China\textsuperscript{2}\\
\{tujh, fanj\}@ruc.edu.cn, \{zeyu.czy, shiyi.zxh, zsq174630, shenxiao.hk, ericzhou.zc\}@alibaba-inc.com
}

\begin{document}
%
\maketitle
\begin{abstract}

The goal of expressive Text-to-speech (TTS) is to synthesize natural speech with desired content, prosody, emotion, or timbre, in high expressiveness. 
Most of previous studies attempt to generate speech from given labels of styles and emotions, which over-simplifies the problem by classifying styles and emotions into a fixed number of pre-defined categories.
In this paper, we introduce a new task setting, \textbf{Contextual TTS} (CTTS). The main idea of CTTS is that how a person speaks depends on the particular context she is in, where the context can typically be represented as text. Thus, in the CTTS task, we propose to utilize such context to guide the speech synthesis process instead of relying on explicit labels of styles and emotions.
To achieve this task, we construct a synthetic dataset and develop an effective framework. 
Experiments show that our framework can generate high-quality expressive speech based on the given context both in synthetic datasets and real-world scenarios.

\end{abstract}
%


\section{Introduction}
\label{sec:intro}
Text-to-speech (TTS) aims to synthesize high-quality and natural speech, where naturalness is determined by the expressiveness of synthesized speech, including content, prosody, emotion, style, etc.
Recently, expressive TTS has received considerable attention,
and most existing methods tend to control synthesized speech through condition embeddings~\cite{gibiansky2017deep, lee2017emotional,DBLP:conf/interspeech/WuWZHSN22} that represent emotions, speakers, styles, etc.
In particular, a commonly-adopted approach to learn condition embeddings is to use explicit labels~\cite{gibiansky2017deep, lee2017emotional}. 
For example, fed with labels representing an emotion category, neural networks will first learn the corresponding emotion embedding, and then use it to further supervise the speech synthesis process. 
%
Another way is to learn condition embeddings from reference speech~\cite{lei2022msemotts,skerry2018towards,wang2018style,DBLP:conf/interspeech/CongYHLX021}. That is, instead of specifying a specific controlling label, it inherits the characteristics of the reference speech. 
However, 
these existing solutions
may incur expressive TTS to be rather inflexible. On the one hand, explicit labeling is costly and may be difficult, because characteristics like styles and emotions are not easy to be categorized.
On the other hand, it may be hard to find appropriate reference speech for the speech to be synthesized. 

In this paper, inspired by the idea that how a speaker says depends on the particular context she is in, we propose to use the \emph{context}, which is a more flexible signal, to guide speech synthesis. Specifically, the tone and mood one uses to say a word depend on the scenes she is experiencing, including when, where, what, etc. For example, under the scenes ``\textit{John hits Alice}'', Alice is expected to speak the text ``\textit{Go away!}'' in an angry tone. Therefore, we introduce a new task setting for expressive TTS, i.e., \textbf{Contextual TTS}~(CTTS), where we describe the scenes in text, then learn the corresponding emotion, personality and other information directly from the contextual text and use it to control speech synthesis. CTTS has high practical value, such as audio book in accordance with the context, or an expressive voice assistant. 

The challenges faced by CTTS fall in text understanding and speech synthesis.
To achieve the CTTS task, we propose a framework including three parts, namely data construction, model building and training strategy. 
Firstly, we propose to construct a new dataset for CTTS based on an existing open-source expressive TTS dataset and an emotional text dataset. The obtained CTTS data has the inputs of both context for semantic understanding and content text for speech synthesis, and produces the corresponding speech of the content text. Secondly, we build an encoder-decoder TTS model for speech synthesis. Lastly, we introduce a three-stage training strategy to enhance the abilities of text understanding and speech synthesis. 
Experiments show that, our model trained on the synthetic dataset can perform well in both synthetic test set and zero-shot real-world data.

Our notable contributions are summarized as follows:
\begin{itemize}
 \item We introduce a new task setting CTTS. To our knowledge, it is the first time using semantic context to control expressive TTS, which is more flexible and generic than previous expressive TTS settings. 
\item We construct a new dataset and develop an effective framework for CTTS\footnote{Our code is based on OFA~\cite{wang2022unifying} and will be released here if the paper is accepted. }, which is simple and efficient.
\item Our experiments show that the speech synthesized by CTTS can ensure accuracy, naturalness and expressiveness both in constructed and zero-shot real-world settings.
\end{itemize}

\section{Methodology}
\label{sec:framework}


The objective of our proposed CTTS task is to extract the emotional, contextual, personal information from the context to guide the speech synthesis. For example, under the context ``\textit{John hit Alice, and Alice said:}'', CTTS is expected to spoken the text ``\textit{Go away!}'' in Alice voice and an angry tone.
In this section, we propose a framework for the CTTS task, which consists of three main parts, namely data construction, model building and training strategy.

\begin{figure}
  \centering
  \includegraphics[width=0.9\linewidth]{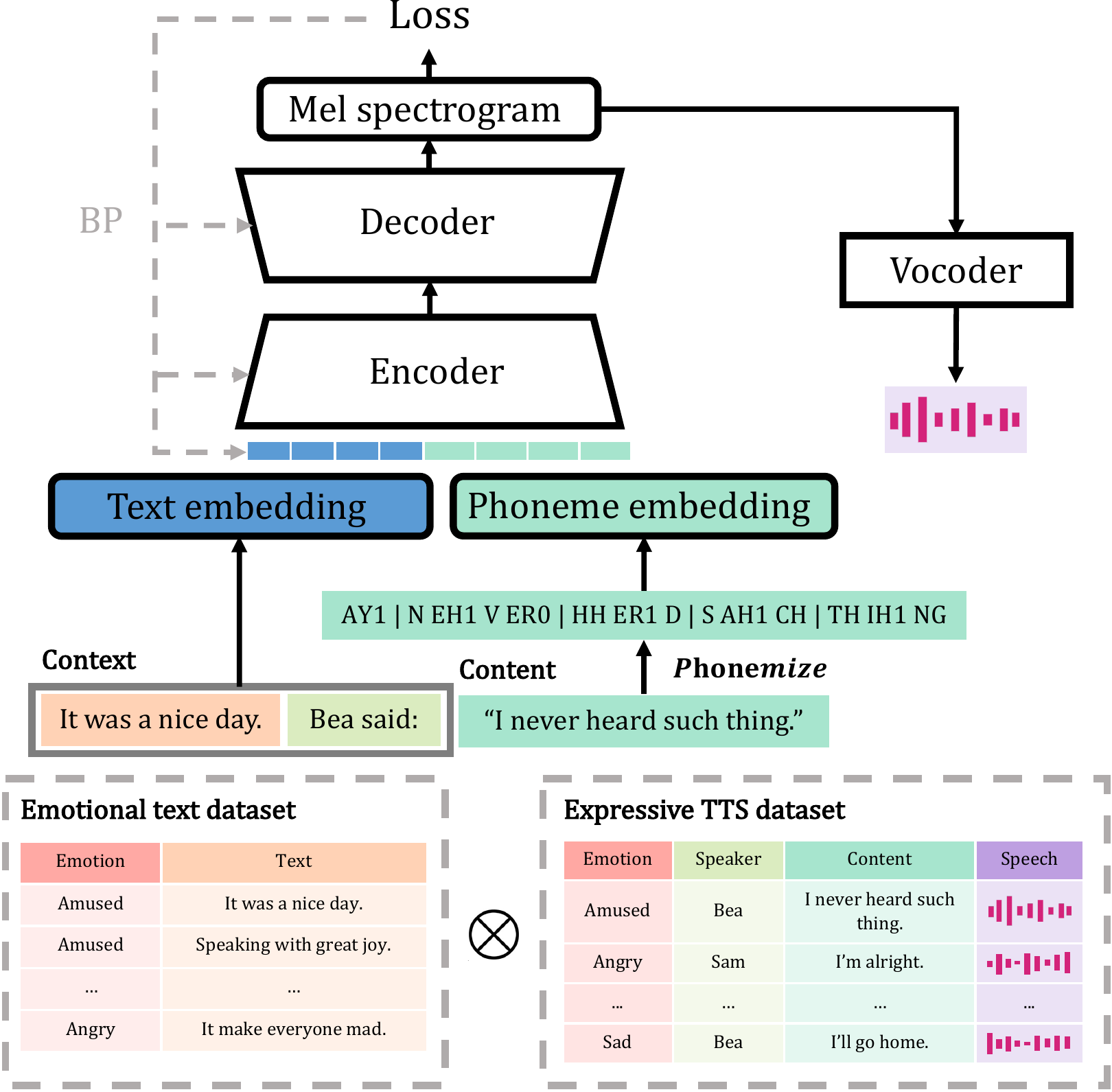}
  \caption{
  The framework of CTTS. The encoder-decoder based model takes context and content as input and outputs mel spectrogram for the content, the pre-trained vocoder converts the mel spectrogram to the speech. The inputted data is constructed by merging two kinds of public datasets, which is described in Section~\ref{subsec:data_construct}.
}
\label{fig:fr}
\end{figure}

\subsection{CTTS Data Construction}
\label{subsec:data_construct}
The training data for CTTS has the following requirements:
(1) the inputs contain both context and content for speech synthesis, and (2) the outputs contain corresponding speech of the content. To fulfill the requirements,
we construct a new dataset by joining an existing open-source expressive TTS dataset and an emotional text dataset, where emotional labels play a role of join key that connects the two datasets.

Specifically, as shown in Figure \ref{fig:fr}, each sample in the expressive TTS dataset contains the speech and its corresponding text (named as content) and other expressive characteristics such as emotion, speaker name, etc, while each sample in the emotional text dataset has text and emotion. For a sample in the expressive TTS dataset, we randomly select a sample with similar emotion in the emotional text dataset. The context is the combination of text in the emotional text dataset and other characteristics in the expressive TTS dataset by a template. 
For example, in Figure \ref{fig:fr}, the context is ``\textit{It’s a nice day. Bea said:}'', which is combined by ``\textit{It’s a nice day}'' and the speaker name ``\textit{Bea}''. 
The content and the output speech are the same as those in the expressive TTS dataset.

Note that we use this synthetic CTTS dataset to guide model to learn the interaction between context and content, and this ability can be generalized to out-of-domain datasets. We evaluate real samples from a popular novel ``\textit{Love in the Time of Cholera}'' and common samples by handwriting in Section~\ref{subsec:zeroshot}.

\subsection{Model Architecture}
\label{subsec:encoder_decoder}
We introduce a model based on the encoder-decoder architecture to predict mel spectrograms for the content considering the context. The predicted mel spectrograms are then synthesized to the speech via a HiFi-GAN vocoder~\cite{kong2020hifi}. The model consists of a context-controlled encoder and an attention-based speech decoder, as discussed below.
\subsubsection{Context-controlled Encoder}
\label{subsec:text_encoder}


The inputs to the encoder include context $c$ and the content $x$.
The context-controlled encoder is expected not only to model the content information of $x$, but also to learn the context information in $c$. 
The inputs are organized as follows. First, we phonemize $x$ to phoneme tokens $p$. Then the text embedding is employed to embed context $c$ into a sequence of vectors $\mathbf{C} = [\mathbf{c_1}, \mathbf{c_2}, ... ] $ while the phone embedding embeds content phonemes $p$ into $\mathbf{P} = [\mathbf{p_1}, \mathbf{p_2}, ... ] $. The encoder input is the concatenation of $\mathbf{C}$ and $\mathbf{P}$. 

The encoder is a multi-layer Transformer. The output of encoder $\mathbf{H}$ is computed:
\begin{equation}
    \mathbf{H} = Encoder([\mathbf{C}, \mathbf{P}]), ~
\end{equation}
$\mathbf{H}$ is supposed to obtain the overall information of emotion, personality, scene and content from $c$ and $x$.

\subsubsection{Attention-based Speech Decoder}
\label{subsec:audio_decoder}
We use the decoder similar to Tacotron2~\cite{shen2018natural} but replace the 2-layer RNN with Transformer. 
The decoder generates stop probability $\mathbf{s}$ and mel spectrograms $\hat{\mathbf{M}} = [\hat{\mathbf{m}}_{1}, \hat{\mathbf{m}}_{1}, ..., \hat{\mathbf{m}}_{M}]$ frame by frame based on the output of encoder $\mathbf{H}$ in an auto-regressive manner, which could be written as,
\begin{equation}
    \hat{\mathbf{M}}, \mathbf{s}  = Decoder(\mathbf{H}),~
\end{equation}
where $M$ is the length of mel spectrograms $\hat{\mathbf{M}}$. 
Finally, we adopt a well-pretrained vocoder HiFi-GAN~\cite{kong2020hifi} to transform $\hat{\mathbf{M}}$ to speech.

\subsection{Training Process}
\label{subsec:training_process}


As the main abilities of our model could be summarized as three parts, speech synthesis, context understanding, and controlling from context to speech, we propose a three-stage training strategy for the CTTS model.

Firstly, to learn the ability of speech synthesis, we pre-train our model using the traditional large TTS dataset LJSpeech~\cite{ljspeech17}. At this stage, we set the controlling context $c$ as blank, and pre-train the phoneme embedding, encoder, and decoder of the CTTS model. The pre-trained model is named as \textbf{M-TTS} in Section \ref{sec:exp}.
Secondly, for better understanding of the context, we propose to improve the ability from large-scale text data. We propose to take advantage of the pre-trained OFA model~\cite{wang2022unifying}, which is trained by around 140G unlabeled text dataset Pile~\cite{DBLP:journals/corr/abs-2101-00027}. Specifically, we employ the well-pre-trained text embedding of OFA to initialize the CTTS model text embedding. 
Finally, based on the pre-trained CTTS model, we use the synthetic CTTS dataset in Section \ref{subsec:data_construct} for training.
Similar to Tacotron2~\cite{shen2018natural}, the training loss includes the mean square error~(MSE) of mel spectrograms and cross-entropy loss on the stop probability of speech synthesis. 


\section{Experiments}
\label{sec:exp}
This section mainly discusses two questions: (1) whether our model could synthesize comparable speech quality as previous TTS settings (Section \ref{subsec:res}); (2) how our model performs in reality scenario, such as out-of-domain novel dubbing, without any labels or reference audios guided (Section \ref{subsec:zeroshot}).

\subsection{Experimental Setup}
\label{subsec:setting}


\begin{table*}[!t]
    \centering
    \caption{Data item examples: \{\textbf{Context}\} is sampled from Yelp, and \{\textbf{Speaker}, \textbf{Content}, \textbf{Speech}\} is from EmoV-DB. They are combined by the same emotional label.} 
    \label{tbl:data}
	\vspace{0.1em}
    \resizebox{\textwidth}{!}{
    \begin{tabular}{|c|c|c|c|}
        \hline
        \textbf{Context} & \textbf{Speaker} & \textbf{Content} & \textbf{Speech} \\
        \hline 
            \hline
            extremely attentive and genuinely a good person. & \multirow{3}{*}{Bee} & \multirow{3}{*}{It was a large canoe.} & \multirow{3}{*}{Amused speech} \\
            \cline{1-1}
            it 's good solid food. & & & \\
            \cline{1-1}
            great place to have some fresh and delicious donuts. & & & \\
        \hline \hline
            it 's just too expensive for what you get. & \multirow{3}{*}{Sam} & \multirow{3}{*}{The men stared into each other's face.} & \multirow{3}{*}{Angry speech} \\
            \cline{1-1}
            needless to say , i will not be returning to this place ever again. & & & \\
            \cline{1-1}
            the only thing i was offered was a free dessert. & & & \\

        \hline \hline
    \end{tabular} 
}
\end{table*}



\noindent
\textbf{Datasets}: 
The CTTS dataset is formed as described in Section \ref{subsec:data_construct}. 
We use the Emotional Voices Database (EmoV-DB)~\cite{adigwe2018emotional} as the experssive TTS dataset and the Yelp Review Dataset (Yelp)~\cite{li2018delete} as the emotional text dataset.
EmoV-DB is a widely used emotional voice dataset that reads transcripts from the CMU Arctic Database~\cite{kominek2004cmu}. For each transcript (\ie content), EmoV-DB contains voices of five emotions from four speakers (two male and two female).
Yelp is a natural language text dataset with obvious emotions, including positive and negative emotions. 
Table~\ref{tbl:data} presents some examples of the reorganized CTTS dataset.
Overall, the CTTS dataset contains 5912 data samples crossing four speakers and two emotions.
Similar to~\cite{kreuk2021textless}, we generate train/valid/test sets with the ratio of 90/5/5, and report performance on test set.


\noindent
\textbf{Evaluation Metrics}: 
The Mean-Opinion-Score (MOS) metric is used to evaluate the emotional similarity between synthetic speech and reference speech. MOS considers speech fluency, prosodic naturalness, emotional correctness and other overall feeling, ranging from 1 to 5 scores. The higher the score, the better the speech quality. 
We release the evaluation task to 21 native English speakers. Each speech is scored by 7 people, and the average value is reported.

To evaluate whether the synthesized speech preserves the lexical content, we use Automatic Speech Recognition (ASR) to evaluate correctness of speech content. We report Word Error Rate (WER) between input content text and the generated text extracted by a pre-trained SOTA ASR system\footnote{We use a LARGE wav2vec 2.0 \cite{baevski2020wav2vec} pretrained and fine-tuned on 960 hours of Librispeech \cite{panayotov2015librispeech} speech audio.}.

\noindent
\textbf{Compared Methods}: 
We present all the models compared in our experiments as follows.

\textbf{(1) M-CTTS}: our proposed model for CTTS task.

\textbf{(2) M-TTS}: pre-training TTS model without any semantic context or label.

\textbf{(3) M-LTTS}: using fixed labels (emotion and speaker) instead of semantic context to control the speech synthesis, which is the traditional setting of expressive TTS.

\textbf{(4) M-CTTS-NT}: M-CTTS without using pre-trained text embedding.

\noindent
\textbf{Implementation Details}: 
We implement the model based on OFA-\textsc{BASE}~\cite{wang2022unifying}. The encoder has 6 transformer layers with 768 hidden embedding size, and the decoder has 6 auto-regressive transformer layers. 
We set the maximum sequence length as 512, the learning rate as 1e-3, batch size as 8 and epoch as 800 for all methods.  
 We use the HiFi-GAN \emph{V}1\footnote{https://github.com/jik876/hifi-gan} pre-trained with LJSpeech~\cite{ljspeech17} as the vocoder.  All the experiments are implemented with 24 NVIDIA TESLA V100 32GB GPUs.

\subsection{Results Analysis}
\label{subsec:res}

In this section, we report and discuss the performances of different methods. All of our speeches are formally displayed on this website\footnote{\url{https://ofa-sys.github.io/Demo_CTTS/}}.

\begin{figure}
  \centering
  \includegraphics[width=0.9\linewidth]{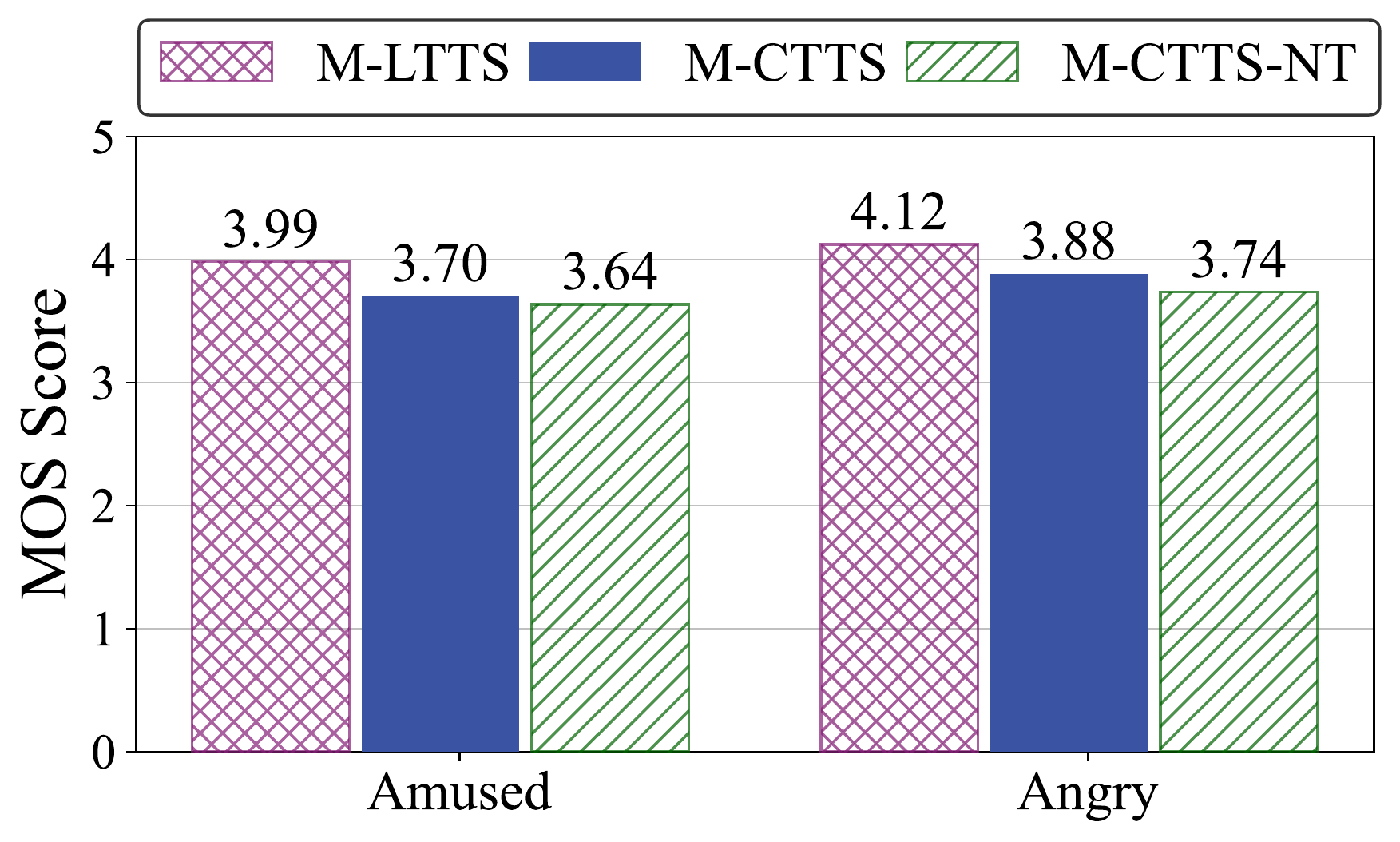}
  \caption{The emotional similarity Mean-Opinion-Score (MOS) results. The two sets of amused and angry are corresponding to two types of semantic contexts with different emotional tendencies. 
  }
\label{fig:mos}
\end{figure}

Figure~\ref{fig:mos} shows the overall MOS results of the models. 
For two types of emotional speeches, M-CTTS generates high scores of 3.70 and 3.88 respectively, which means that the overall styles, prosodies, and emotions of the speeches we synthesized are mostly the same as those of the real referenced speeches.
M-LTTS achieves the best performance, which is reasonable because we provide accurate labels in the synthetic test set. However, compared with M-LTTS, M-CTTS has the advantage of text understanding and can handle out-of-domain contexts, which will be reported in Section~\ref{subsec:zeroshot}. 
Compared with M-CTTS-NT, the superior performance of M-CTTS shows that the text understanding ability from pre-trained language model is helpful for expressive speech synthesis in the CTTS task.

Table~\ref{tbl:asr} shows the experimental results with respect to WER. 
The results show that the M-TTS model has the lowest WER (0.0734) which means that the pre-trained TTS model has strong speech synthesis ability to ensure the accuracy of the content. Recall that the pre-trained TTS model does not use any expressive speech dataset for training, and thus the synthetic speeches only have a flat tone without emotion. 
The WER scores of the first four emotional speeches are higher than that of pre-trained TTS. This is because emotional speeches contain more emotional connectives, such as laughter and sigh, which lead to greater differences between the recognized text by the ASR model and the standard text.
Compared with Ref* (0.1401), M-CTTS (0.1342) and M-LTTS (0.1122) achieve slightly better WER scores, which means the synthesis qualities of M-CTTS and M-LTTS are reliable. The inferior performance of M-CTTS-NT (0.2186) shows that the text understanding ability of M-CTTS has a great influence on the quality of speech synthesis.



\subsection{Zero-shot Cases in Reality Scenario}
\label{subsec:zeroshot}
Although M-CTTS is trained on synthetic datasets, it also performs well in real scenarios. 
To show this, we evaluate out-of-domain examples where no explicit labels or reference speech could be found. We select some semantic context and content pair from a popular novel ``\textit{Love in the Time of Cholera}'' and manually write some common descriptions about emotion, weather and personalities as controlling context. 


The speeches synthesized by our model can be download from the website\footnote{\url{https://ofa-sys.github.io/Demo_CTTS/}}. 
The results show that in these out-of-domain and real-world scenarios, M-CTTS can still synthesize expressive speeches. 

Recall that in our constructed CTTS dataset, Yelp dataset only contains a small amount of text data related to reviews, and the number of speeches of EmoV-DB is very small. The success on out-of-domain examples indicates that M-CTTS can effectively generalize the text understanding ability of the pre-trained language model (\ie OFA), retain the speech synthesis ability of the pre-trained TTS model, and achieve fine-tuning the prosody, emotion, etc of the synthetic speech, thus improving the speech expression ability.

\begin{table}[!t]
    \centering
    \caption{The WER results for our proposed M-CTTS and baselines. \textbf{Ref*} denotes ground truth speech.
    } 
    \label{tbl:asr}
	\vspace{0.1em}
    \resizebox{\columnwidth}{!}{
    \begin{tabular}{c|c|c|c|c|c}
        \hline
         Metrics &\textbf{Ref*} & \textbf{M-CTTS} & \textbf{M-LTTS} & \textbf{M-CTTS-NT} & \textbf{M-TTS} \\
        \hline 
            \hline
           WER & 0.1401 &0.1342 & 0.1122& 0.2186& 0.0734\\
           \hline
        \hline 
    \end{tabular} 
}
\end{table}
\section{Conclusion}
\label{sec:conclusion}
In this paper, we have introduced a new task CTTS setting using context to control expressive speech synthesis, which is more flexible and has better generalization capability than previous settings. The context is represented as text here.
To achieve this task, we constructed a CTTS dataset based on emotional text datasets and expressive TTS datasets. 
The model we proposed for the task supports the simultaneous understanding of context semantics and speech content text, and the generation of expressive speech. 
The experiments show that our model can synthesize correct, natural and expressive speeches from contextual information in both synthetic dataset and real-world out-of-domain scenarios.



In the future, we will collect more real-world data to extend CTTS dataset for training. Moreover, the context can be represented as not only text but also other modalities, such as video, audio, and image.





\vfill\pagebreak

\bibliographystyle{IEEEbib}
\bibliography{ref/ref.bib}

\end{document}